# Coriolis Force Induced Quantum Hall Effect for Phonons

Yao-Ting Wang[1], Pi-Gang Luan[2], and Shuang Zhang[1*]

[1]*School of Physics and Astronomy, University of Birmingham, Birmingham B15 2TT, United Kingdom*

[2]*Department of Optics and Photonics, National Central University, Jhongli 32001, Taiwan*

**A two-dimensional mass-spring system with Honeycomb lattice for mimicking phononic quantum Hall effect is proposed. Its band structure shows the existence of Dirac cones and unconventional edge states that is similar to the vibrational modes in graphene. Interestingly, as the system is placed on a constantly rotational coordinate system, the *Coriolis force* resulted from the non-inertial reference frame provides a possibility to break the time-reversal symmetry. Thus, caused from topologically non-trivial band gaps, phononic edge states are present between bands, which are verified by the calculation of Chern numbers for corresponding bands.**

Over the past few decades, the discovery of matter with topological nature, such as quantum Hall effect and topological insulators, has attracted much attention on these novel phenomena [1-5]. Topological protection, the immunity against scattering of matter waves by disorder, is a main property existing in these kinds of materials. The scattering-immunity states probably are the most impressive concepts of materials because this robustness of electronic waves was previously only known in superconductors. Theoretical studies show that breaking time-reversal symmetry plays a crucial role to open a non-trivial frequency gap, which is an essential condition to bring about the topologically protected edge states [6]. The condition of time-reversal symmetry breaking can be fulfilled through applying an external magnetic field, or the intrinsic spin-orbit coupling of the material itself. Recently, the thriving progress of topological properties in condensed matter has also led to a great interest in discovering similar phenomenon in other physics systems. S. Raghu and F. D. M. Haldane [7-8] proposed that the optical analogue of quantum Hall effect could be achieved by using periodically arranged gyromagnetic rods. Following this idea, many research groups have realized topologically protected one-way edge states theoretically [8-10] and experimentally [11]. Also, by employing periodically arranged meta-crystals and helical arrays two approaches of photonic analogue are proposed for realizing photonic topological insulator which breaks time-reversal symmetry for each spin but preserved it in the whole system [12-13]. Nevertheless, only few works focus on the analogues of quantum spin Hall effect in acoustic or phononic systems [14-17]. In this Letter, we theoretically show that a phononic crystal in a rotating frame can possess the similar features of quantum-Hall-effect edge states.

We consider a two-dimensional spring-mass system with honeycomb lattice in the non-rotating lab frame. The system consists of a series of honeycomb-arranged particles with mass $M$ and springs with elastic constant $C$, as shown in Fig. 1a. All particles are considered as a rigid body, and the mass of springs is omitted. According to the above descriptions and the labelling approach [18] shown in Fig. 1b, if we assume the general form $\xi(m,n) = \xi p^m s^n e^{-i\omega t}$, $\boldsymbol{\eta}(m,n) = \boldsymbol{\eta} p^m s^n e^{-i\omega t}$, the systematic equations of dispersion relation are expressed as [19]

$$\omega^2 \boldsymbol{\xi} = \omega_0^2 \left[ \hat{\mathbf{R}}_1 \hat{\mathbf{R}}_1 \cdot (\boldsymbol{\xi} - \boldsymbol{\eta}) + \hat{\mathbf{R}}_2 \hat{\mathbf{R}}_2 \cdot (\boldsymbol{\xi} - p^* \boldsymbol{\eta}) + \hat{\mathbf{R}}_3 \hat{\mathbf{R}}_3 \cdot (\boldsymbol{\xi} - s^* \boldsymbol{\eta}) \right], \tag{1-a}$$

$$\omega^2 \boldsymbol{\eta} = \omega_0^2 \left[ \hat{\mathbf{R}}_1 \hat{\mathbf{R}}_1 \cdot (\boldsymbol{\eta} - \boldsymbol{\xi}) + \hat{\mathbf{R}}_2 \hat{\mathbf{R}}_2 \cdot (\boldsymbol{\eta} - p\boldsymbol{\xi}) + \hat{\mathbf{R}}_3 \hat{\mathbf{R}}_3 \cdot (\boldsymbol{\eta} - s\boldsymbol{\xi}) \right]. \tag{1-b}$$

where $p = e^{i\mathbf{k}\cdot\mathbf{a}_1}$, $s = e^{i\mathbf{k}\cdot\mathbf{a}_2}$, and $\omega_0^2 = C/M$ is characteristic frequency of spring mass system. By solving the eigenvalue problem $\mathbf{H}\mathbf{x} = \omega^2 \mathbf{x}$ given by (1), we obtain four solutions, two of them are 0 and $\sqrt{3}\omega_0$, while the other two are described by the following dispersion relation,

$$\omega = \omega_0 \sqrt{\frac{3}{2} \mp \frac{1}{2}\sqrt{3 + 2\{\cos(\mathbf{k}\cdot\mathbf{a}_1) + \cos(\mathbf{k}\cdot\mathbf{a}_2) + \cos[\mathbf{k}\cdot(\mathbf{a}_1 - \mathbf{a}_2)]\}}}. \tag{2}$$

The band structure described by equation (2) is shown in Fig. 1c. There exist four bands, with the two flat bands of transverse modes at zero frequency and $\sqrt{3}\omega_0$, and two dispersive longitudinal modes. These two flat bands arise from the assumption of negligible transverse restoring force of the springs. If the transverse restoring force is present in a similar way as in graphene, both the optical and acoustic transverse mode would become dispersive as shown in [20]. Six linear degeneracies are seen in the vicinity of K points around the whole Brillouin zone. By expanding equation (2) to the second order at K point $\left(4\pi/3\sqrt{3}a,0\right)$, we show that $\omega=\omega_0\left(1\pm|\delta\mathbf{k}|a/4\right)$ is a linear function with respect to **k** vector. Fig. 1d indicates the zoomed-in plot nearby K point. These degeneracies, referred to as Dirac points, provide the chance to exhibit topological edge modes being known as the quantum spin Hall effect when the time reversal symmetry is broken.

To investigate the phononic edge states, we study an infinitely long ribbon along *x* direction, but a finite width consisted of *N* unit cells in *y* direction. The band structure of spring-mass ribbon having honeycomb lattice is shown in Fig. 2. Similar to graphene ribbon ending with bearded edges, a flat edge state between two Dirac points is observed in the energy spectrum fig. 2b. Due to existence of the flat energy band, at the corresponding frequency, the particles located on the edge vibrate locally without transporting energy forward. For the bearded edge, band diagram shown in Fig. 2c also gives a flat band throughout the whole **k** space. We note that it differs from conventional bearded flat edge mode for electrons in graphene, which only appears in the region between the two K points. In addition, two unconventional edge states (red curves) are found around the Dirac points in Fig. 2b and 2c. These new states for beaded or zigzag boundary were recently proposed and observed by Yonatan Plotnik *et*. *al*. in a photonic analog system of graphene [21] and also the theoretically investigated for the phonons in real graphene [22]. Fig. 2d and 2e show the vibration amplitudes at points A along the transverse cross-section (N = 20) of spring-mass ribbon. As predicted by energy spectrum, the peaks indicate that the vibration of particles is highly confined to the corresponding edge.

Next we introduce time reversal symmetry breaking by placing the spring mass lattice in a rotational frame. In classical mechanics, objects moving on a rotational reference frame experiences inertial forces - Centrifugal force and Coriolis force, resulting in corrections to the equation of motion. For a particle doing simple harmonic oscillation on a counter-clockwise rotating round plate with a constant angular velocity $\mathbf{\Omega}$ in $z$ direction, the Lagrangian in terms of $r$ and $v$ can be expressed as $L = \left[ Mv^2 - Cr^2 + M(\mathbf{\Omega} \times \mathbf{r})^2 \right]/2 + M\mathbf{v} \cdot (\mathbf{\Omega} \times \mathbf{r})$ [23]. The first two terms denote normal simple harmonic oscillation, the third leading to a correction term of elastic potential energy is the centrifugal force, and the fourth one corresponds to the Coriolis force. Then, letting $\mathbf{A} = (\mathbf{\Omega} \times \mathbf{r})/2$ that serves as a counterpart of Coulomb gauge in quantum physics, Lagrangian can be rewritten as $L = \left[ Mv^2 - (C - M\Omega^2)r^2 \right]/2 + 2M\mathbf{A} \cdot \mathbf{v}$. It is clear that the Lagrangian of rotating reference frame can be analogous to the two dimensional system under a constant magnetic field along normal direction with the corrected elastic constant and relevant 'charge' Q = 2$M$. On the other hand, the presence of centrifugal force will disturb the system and break the translational symmetry of the lattice. However, since the centrifugal force has a quadratic dependence on the angular frequency of the rotation frame, its effect can be neglected for sufficiently small angular frequencies. More detailed discussion of effect of the centrifugal force will be given later. With only the Coriolis force taken into account, the dynamic equations can be written as,

$$\omega^2 \boldsymbol{\xi} + 2i\omega\Omega \hat{\mathbf{z}} \times \boldsymbol{\xi} = \omega_0^2 \left[ \hat{\mathbf{R}}_1 \hat{\mathbf{R}}_1 \cdot (\boldsymbol{\xi} - \boldsymbol{\eta}) + \hat{\mathbf{R}}_2 \hat{\mathbf{R}}_2 \cdot (\boldsymbol{\xi} - p^*\boldsymbol{\eta}) + \hat{\mathbf{R}}_3 \hat{\mathbf{R}}_3 \cdot (\boldsymbol{\xi} - \boldsymbol{\eta}_{n-1}) \right], \quad \text{(3-a)}$$

$$\omega^2\mathbf{\eta}+2i\omega\Omega\hat{\mathbf{z}}\times\mathbf{\eta}=\omega_0^2\left[\hat{\mathbf{R}}_1\hat{\mathbf{R}}_1\cdot(\mathbf{\eta}-\mathbf{\xi})+\hat{\mathbf{R}}_2\hat{\mathbf{R}}_2\cdot(\mathbf{\eta}-p\mathbf{\xi})+\hat{\mathbf{R}}_3\hat{\mathbf{R}}_3\cdot(\mathbf{\eta}-\mathbf{\xi}_{n+1})\right], \quad \text{(3-b)}$$

The band structure can be obtained by solving the eigenvalue equation. Fig. 3a-c show the band structures at three different angular velocities of the rotation frame, $\Omega$ = 1, 4, 12, and 15 [Hz], respectively. At the lowest angular velocity $\Omega$ = 1 [Hz], interestingly, the top and bottom bands evolves from the originally flat bands in the non-rotating frame to have finite bandwidth. There are three band gaps between the four bands are topologically nontrivial because of time-reversal symmetry breaking, and there exist topologically protected edge states within each bandgap, as shown in Fig. 3a. One notes that the three sets of top nontrivial edge states evolve from the flat middle edge state, and the two unconventional edge states in the non-rotating frame.

As the rotating speed increases to 4 [Hz], the band gaps broaden and the dispersion of the edge states appear more linear. When the rotating speed of the frame reaches to about 12 [Hz], the second and third band touch each other and the gap between them is closed. When the rotational speed is further increased, the gap reopens but it is now converted into a trivial one, which does not contain any edge states due to the band crossing, as shown in Fig. 3c. Fig. 4a to 4c show the bearded edge states under different speeds of rotation. Similar to the zigzag edge, the bearded edge states exist in every band gap as the angular velocity is less than 12 [Hz]. Due to band inversion, topological phase transition occurs at $2^{nd}$ and $3^{rd}$ band after reaching to the critical angular frequency.

The trajectories of particles of three edge states within band gaps are shown in Fig. 5. Particles movement follows the red-green-yellow-white color sequence periodically and the same color dots denote positions of particles at the same time. It is obvious to note that three states are localized on the edge. If we further look into their propagating behaviors, for zigzag edge type, there are three distinct types observed in the figure. The 1$^{st}$ edge mode expresses a clockwise rotation of particles propagating along one direction. The 3$^{rd}$ one, however, presents counter-clockwise movement confined to the edge. This anti-symmetric moving state gives rise to the higher operating frequency in the band diagram. The vibration emerging from the central energy gap remains bounded at the edge, but its vibrational expression is linear from other modes. On the other hand, vibrational modes for bearded edge shows different characteristic from zigzag one. As presented in fig. 5b, two edge particles in 1$^{st}$ and 2$^{nd}$ gap vibrate in phase as the clockwise and counter-clockwise rotation, respectively. The edge mode in 3$^{rd}$ gap, however, rotates with 180° phase difference, which leads to the highest frequency in band structure.

We now address the influence of equilibrium position correction affected by centrifugal force. Under the approximation $\Omega^2 \ll \omega_0^2$, the new equilibrium position of each particle is given by

$$\Delta \mathbf{a}_{mn}^{A,B} = \frac{2\Omega^2}{3\omega_0^2}\left(\mathbf{r}_{mn}^{A,B} + \mathbf{r}_0\right), \tag{4}$$

where $\mathbf{r}_{mn}^{A,B}$ is position vectors for particle A and B related to the center of the strip and $\mathbf{r}_0$ is an initial vector with respect to the original point. To estimate the effect of the centrifugal force, we consider a mass-spring ribbon with characteristic frequency $\omega_0 = 20$ Hz located on a round plate, which has a width of ten periods (N = 10), a length that is two times of the width, and a lattice constant $a = \sqrt{3}$ cm. The center of ribbon is located at the center of round plate ($\mathbf{r}_0 = 0$). For angular velocity $\mathbf{\Omega} = 2$ [Hz], the farthest particle from center is shifted from its original position by 0.129 cm due to the centrifugal force. It has been shown that the relation between disorder and deviation of band gap is linked by an approximate formula

$\delta^2 \sim \Delta\omega/3\omega_{cen}$ or $\Delta\omega \sim 0.017\omega_{cen}$ [24], where $\delta$ is the deviation parameter defined as $\Delta a/a$, $\Delta\omega$ is the change of bandgap, and $\omega_{cen}$ gives the central frequency of a certain gap. The gap widths for each band are 2.122, 4.010, and 2.122 Hz, respectively. It is obvious that the change of bandgap is so small that it can be neglected for this particular example. Hence, as long as the angular velocity is much less than characteristic frequency, the influence of equilibrium position shift is insignificant to the band structure. For operation at higher angular velocity, one can still sets all particles at corrected positions in advance such that when round plate rotate at certain angular velocity, all particles move back to the original ordered locations.

To confirm the topological order of edge states, we have numerically evaluated Chern number for each band. The formula of Chern number is expressed as $C_n = (1/2\pi)\int_{BZ} d^2\mathbf{k}\cdot\mathbf{F}_n$, where $\mathbf{F}_n = i\langle\nabla_\mathbf{k} n|\times|\nabla_\mathbf{k} n\rangle$ is the Berry curvature for the $n$th band, and the integration is taken throughout the first Brillouin zone. The result shows that, at $\Omega = 4$ Hz, Chern numbers from the lowest to the highest band are given as $\{-1,0,0,1\}$. It shows edge states in bands are topologically protected in association with net Chern number $\Delta C = 2$. When $\Omega$ reaches 12 Hz, topological transition occurs so that the Chern numbers for bands become $\{-1,1,-1,1\}$. The system now exhibits topologically non-trivial phase difference $\Delta C = 2$ only in the $1^{st} - 2^{nd}$ and $3^{rd} - 4^{th}$ bands. The central band gap now becomes nontrivial, and there exist no scattering immune edge states. The Chern number calculation agrees perfectly with the results shown in fig. 3 and 4.

In summary, we have introduced a simple phononic system - a honeycomb lattice plane consisting of rigid bodies and soft springs, which, when positioned in a rotation frame, can have non-trivial frequency gap that gives rise to appearance of topologically protected edge states. These edge states should be observed experimentally at low-frequency oscillation.

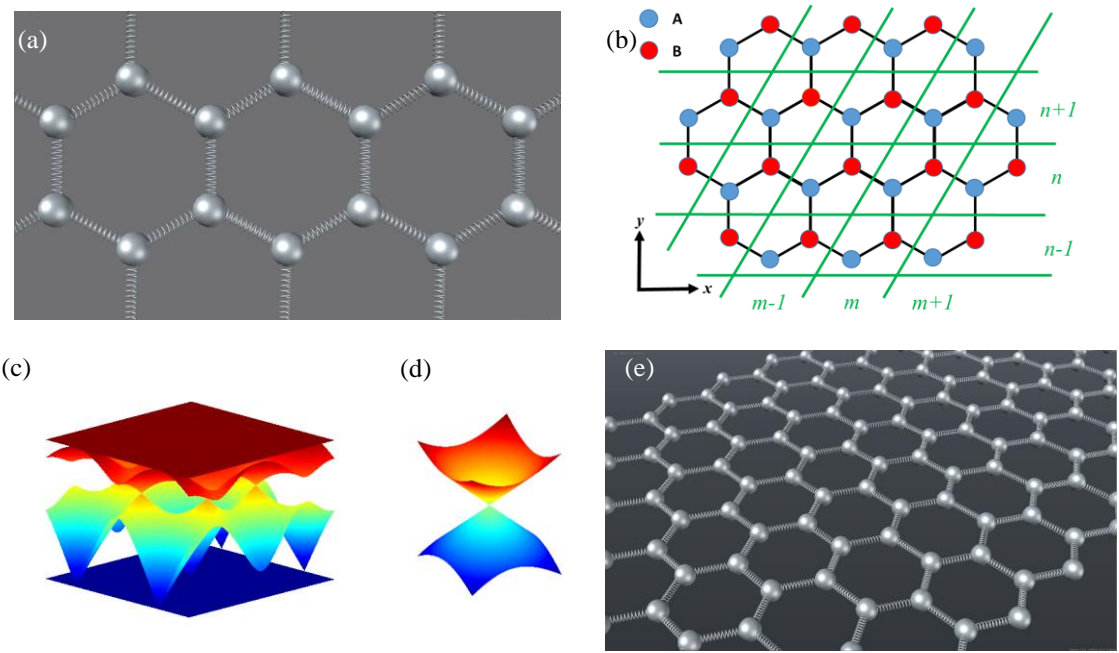

Figure 1. (a) Top view of our model made by soft springs and rigid particles. (b) Labelling the two-dimensional spring-mass model utilizing integer *m* and *n*. (c) The band structure of two-dimensional mass-spring model with honeycomb lattice and (d) a zoomed-in figure of Dirac cone in the vicinity of a certain K point with mass and elastic constant equals to ten grams and 4 N/m, respectively. (e) A three-dimensional plot of the semi-infinite mass-spring ribbon ending with zigzag edges.

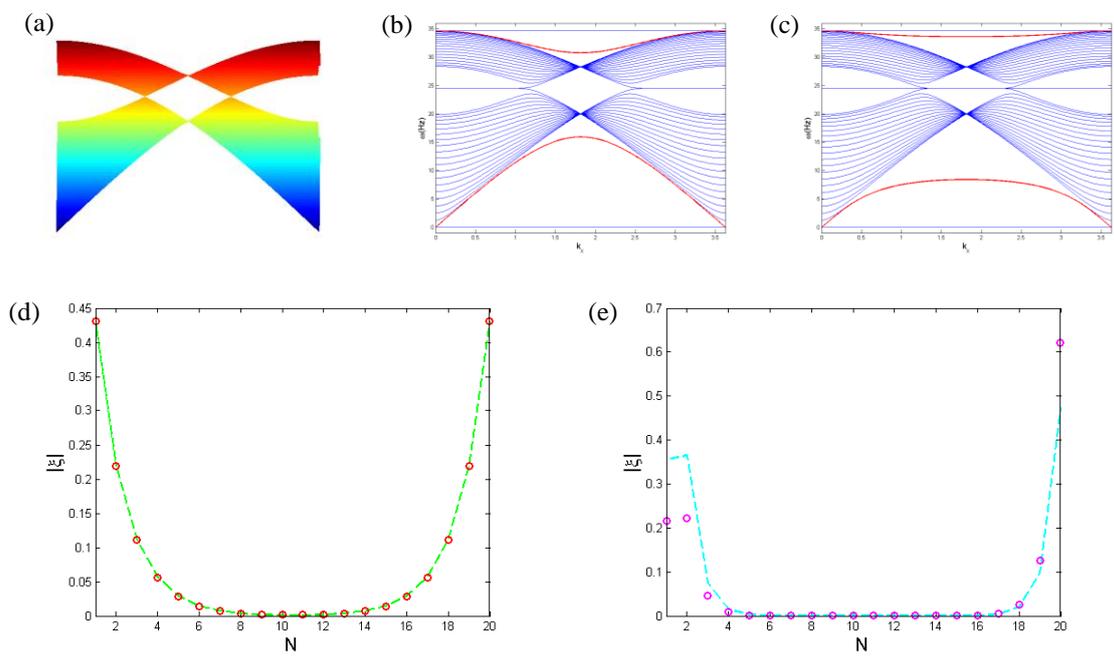

Figure 2. (a) The side view of three-dimensional band structure compares with (b) the one-dimensional band diagram calculated from zigzag spring-mass ribbon. (c) One-dimensional band diagram for bearded edges. (d) Absolute value of amplitudes for top and bottom edge modes, the green dashed line corresponds to the top edge state and the red circles are to the opposite one. (e) Its absolute value of amplitudes of fig. 2c for both edge states.

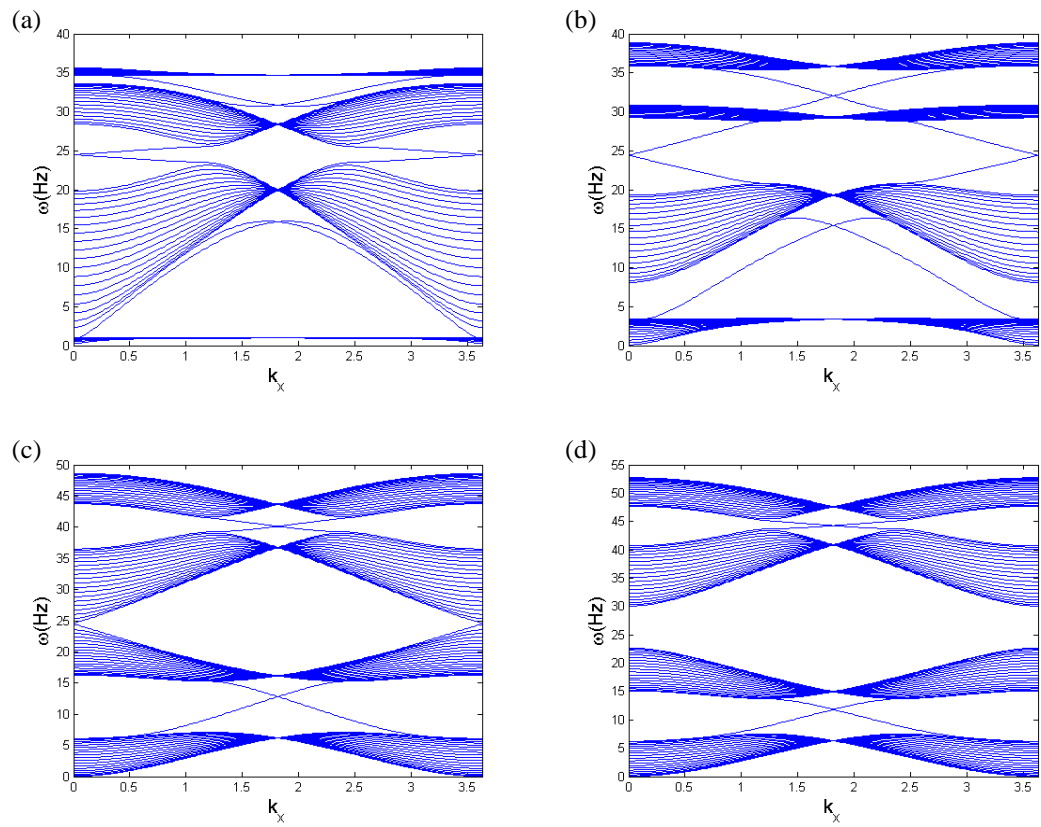

Figure 3. One-dimensional band structure for a mass-spring ribbon end with zigzag boundary applied constantly angular velocity (a) $\Omega = 1$, (b) $\Omega = 4$, (c) $\Omega = 12$, and (d) $\Omega = 15$ Hertz.

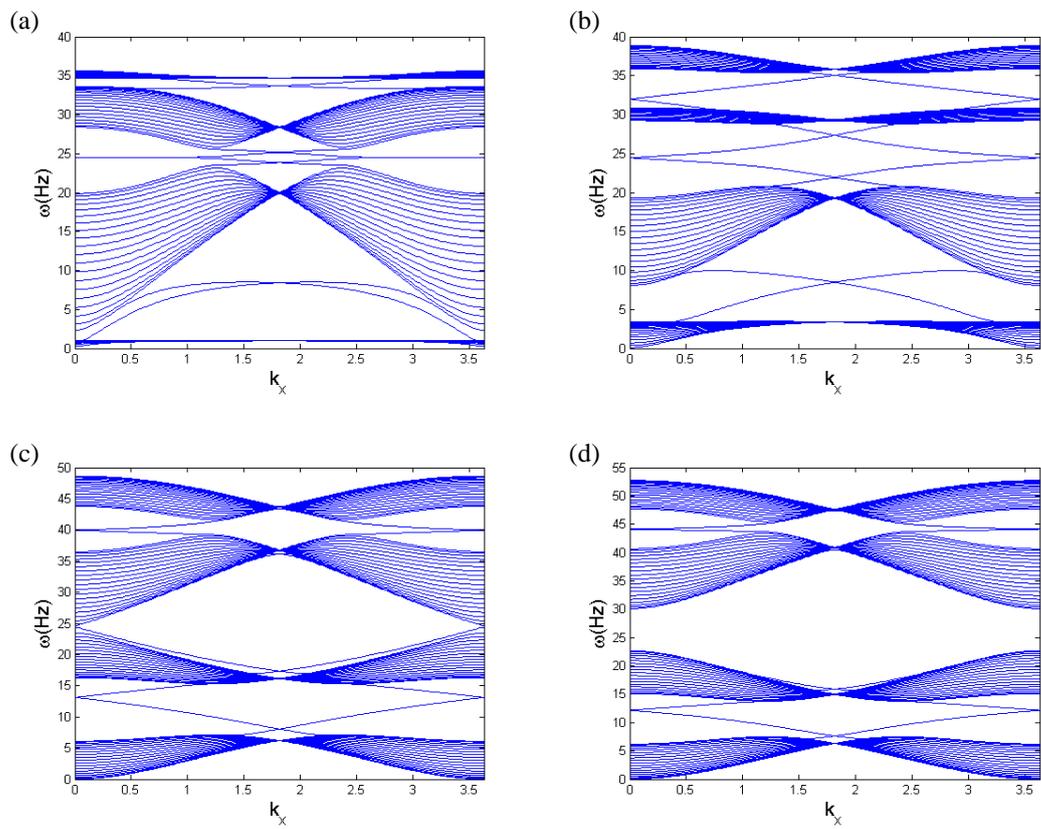

Figure 4. Frequency diagram in one dimension for a mass-spring ribbon end with bearded boundary applied constantly angular velocity at (a) $\Omega = 1$, (b) $\Omega = 4$, (c) $\Omega = 12$, and (d) $\Omega = 15$ Hertz.

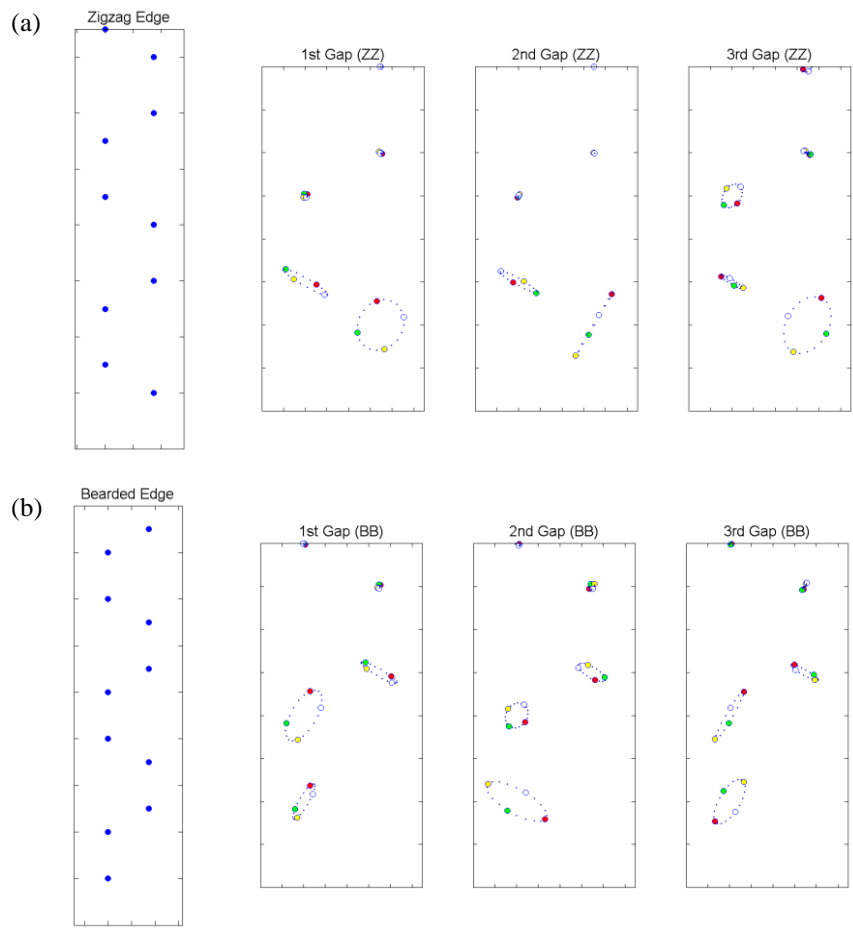

Figure 5: The vibrational trajectories of a spring-mass strip ended with (a) zigzag and (b) bearded boundary. Particles move by following the red-green-yellow-white color sequence. The same color dots mean position of particles at the same time.